\documentclass[prd,preprint,superscriptaddress,showpacs,byrevtex]{revtex4}
\usepackage{bm}
\def\fsl#1{\setbox0=\hbox{$#1$}           % set a box for #1
   \dimen0=\wd0                                 % and get its size
   \setbox1=\hbox{/} \dimen1=\wd1               % get size of /
   \ifdim\dimen0>\dimen1                        % #1 is bigger
      \rlap{\hbox to \dimen0{\hfil/\hfil}}      % so center / in box
      #1                                        % and print #1
   \else                                        % / is bigger
      \rlap{\hbox to \dimen1{\hfil$#1$\hfil}}   % so center #1
      /                                         % and print /
   \fi}                                         %
\newcommand{\be}{\begin{equation}}
\newcommand{\ee}{\end{equation}}
\newcommand{\bea}{\begin{eqnarray}}
\newcommand{\eea}{\end{eqnarray}}
\newcommand{\beq}{\begin{equation}}
\newcommand{\eeq}{\end{equation}}
\newcommand{\beqs}{\begin{eqnarray}}
\newcommand{\eeqs}{\end{eqnarray}}

\newcommand{\dslash}{D\hspace{-0.067in}\slash}
\begin{document}
\title{ Proof of Factorization Using Background Field Method of QCD }
\author{Gouranga C Nayak } \email{nayak@physics.arizona.edu}
\affiliation{ Department of Physics, University of Arizona, Tucson, AZ 85721, USA}
\begin{abstract}
Factorization theorem plays the central role at high energy colliders to study standard model and beyond
standard model physics. The proof of factorization theorem is given by Collins, Soper and Sterman to all orders
in perturbation theory by using diagrammatic approach. One might wonder if one can obtain the proof of factorization
theorem through symmetry considerations at the lagrangian level. In this paper we provide such a proof.
\end{abstract}
\maketitle
\pagestyle{plain}
\pagenumbering{arabic}
Factorization theorem plays the central role at high energy colliders to study standard model and/or beyond
standard model physics. The proof of the factorization theorem is very non-trivial and is given by Collins,
Soper and Sterman by using diagrammatic approach to all orders in perturbation theory \cite{sterman,collins}
(see also \cite{cs2,cs3}). Factorization refers to separation of short-distance from long-distance effects in field theory.
Using the factorization theorem the
hadron production cross section at high energy hadronic colliders can be calculated by using the well known formula
\bea
\sigma (AB \rightarrow H+X) ~=~ \sum_{a,b,c,d}~\int dx_1 ~\int dx_2 ~\int dz~f_{a/A}(x_1,Q^2)~f_{b/B}(x_2,Q^2)~{\hat \sigma}(ab \rightarrow cd)~D_{H/c}(z,Q^2) \nonumber \\
\label{cr1}
\eea
where
\bea
&& f_{q/P}(x,k^2_T)
= \frac{1}{2}~\int dy^- \frac{d^{d-2}y_T}{(2\pi)^{d-1}}  e^{ix{P}^+ y^- + i {k}_T \cdot y_T} \nonumber \\
&&~\frac{1}{2}{\rm tr_{Dirac}}~\frac{1}{3}{\rm tr_{color}}[\gamma^+<P| {\bar \psi}(y^-,y_T) ~\Phi[y^-,y_T] ~\Phi^{-1}[0]~{ \psi}(0)  |P>].
\label{pdf}
\eea
is the parton distribution function and
\bea
&& D_{H/q}(z,P^2_T)
= \frac{1}{2z} \int dx^- \frac{d^{d-2}x_T}{(2\pi)^{d-1}}  e^{i{k}^+ x^- + i {P}_T \cdot x_T/z} \nonumber \\
&&~\frac{1}{2}{\rm tr_{Dirac}}~\frac{1}{3}{\rm tr_{color}}[\gamma^+<0| {\bar \psi}(x^-,x_T) ~\Phi[x^-,x_T]~a^\dagger_H(P^+,0_T)  a_H(P^+,0_T) ~\Phi^{-1}[0]~{ \psi}(0)  |0>].
\label{qvf}
\eea
is the fragmentation function \cite{collins}. In the above expression $a^\dagger_H$ is the
creation operator of a hadron. The Wilson line is given by
\bea
\Phi[x^\mu ]={\cal P}~ {\rm exp}[ig\int_{-\infty}^{0} d\lambda~ h \cdot {\cal A}^a(x^\mu +h^\mu \lambda )~T^a]
\label{wilf}
\eea
where $h^\mu$ is a $x^\mu$ independent four vector.

One might wonder if one can obtain the proof of factorization theorem through symmetry considerations at the
lagrangian level. Such an attempt was made in \cite{tucci} to prove factorization of soft and collinear divergences by
using background field method of QCD \cite{thooft,abbott}. However, the proof presented in \cite{tucci} is incomplete for QCD
although it is complete for QED \cite{nayakqed}. The main difficulty is due to the gauge fixing term in the background field
method of QCD which, unlike QED, depends on the background field \cite{thooft,abbott}. Recently we have derived a gauge fixing identity
\cite{nayakgf} which relates the generating functional in QCD with the generating functional in the background field method of
QCD in pure gauge. In this paper we will provide a proof of the factorization theorem in high energy QCD by using this gauge fixing
identity.

The most crucial statement of factorization theorem of Collins, Soper and Sterman is the appearance of the
Wilson line eq. (\ref{wilf}) in the definition of the structure functions and fragmentation functions
in eqs. (\ref{pdf}) and (\ref{qvf}) which makes them gauge invariant. This Wilson line is responsible for
cancelation of soft and collinear divergences which arise due to presence of loops and/or higher order Feynman
diagrams. This Wilson line can be thought of as a quark or gluon jet propagating in a soft and/or collinear gluon
cloud. One of the ideas to use background field method of QCD is to first show that this soft and collinear
gluon cloud can be represented as a classical background field ${\cal A}_\mu^a(x)$ in an abelian-like pure gauge
\bea
{\cal A}_\mu^a = \partial_\mu \omega^a(x).
\label{apure}
\eea
This can be shown as follows.

The Wilson line eq. (\ref{wilf}) can be written as
\bea
&&~\Phi[x^\mu ]={\cal P}~ {\rm exp}[ig\int_{-\infty}^0 d\lambda~ h \cdot {\cal A}^a(x^\mu +h^\mu \lambda )~T^a] \nonumber \\
&& ={\cal P}~ {\rm exp}[ig\int_{-\infty}^0 d\lambda~ h \cdot e^{\lambda h \cdot \partial} {\cal A}^a(x^\mu )~T^a]
={\cal P}~ {\rm exp}[ig \frac{1}{ h \cdot \partial} h \cdot  {\cal A}^a(x^\mu )~T^a].
\label{wilf1}
\eea
Using the Fourier transformation
\bea
{\cal A}^a_\mu(x) =\int \frac{d^4k}{(2\pi)^4} {\cal A}^a_\mu(k) e^{ik \cdot x}
\label{ft}
\eea
we find from eq. (\ref{wilf1}) the phase factor
\bea
V=g~\omega(k)=ig~\frac{h \cdot {\cal A}(k)}{h \cdot k}
\label{eikonal}
\eea
which is precisely the Eikonal Feynman vertex for soft and collinear divergences.
For example if we choose $h^\mu=n^\mu$, where $n^\mu$ is a fixed lightlike vector having only "+" or "-" component \cite{sterman}
\bea
n^\mu=(n^+,n^-,n_T)=(1,0,0)~~~~~~~~~~~{\rm or}~~~~~~~~~n^\mu=(n^+,n^-,n_T)=(0,1,0)
\label{n}
\eea
we reproduce the divergences due to soft gluons. Similarly, if we choose $h^\mu=n^\mu_B$, where $n^\mu_B$ is a non-light like vector
\bea
n^\mu_B=(n^+_B,n^-_B,0),
\label{nb}
\eea
we reproduce the Feynman rules for the collinear divergences \cite{sterman}. Multiplying a $x^\mu$ independent free vector $h^\mu$ and dividing $h \cdot \partial$ from left in eq. (\ref{apure})
we find
\bea
\omega^a(x) = \frac{1}{h \cdot \partial} h \cdot {\cal A}^a
\label{wilf1a}
\eea
which is exactly the phase factor that appears in the Wilson line in eq. (\ref{wilf1}). This establishes the correspondence between the Wilson
line in eq. (\ref{wilf}) and the classical field in an abelian-like pure gauge as given by eq. (\ref{apure}).

In QCD, the generating functional is given by
\bea
Z_{\rm QCD}[J,\eta,{\bar \eta}]=\int [dQ] [d{\bar \psi}] [d \psi ] ~{\rm det}(\frac{\delta \partial_\mu Q^{\mu a}}{\delta \omega^b})
~e^{i\int d^4x [-\frac{1}{4}{F^a}_{\mu \nu}^2[Q] -\frac{1}{2 \alpha} (\partial_\mu Q^{\mu a})^2+{\bar \psi} \dslash [Q] \psi + J \cdot Q +{\bar \eta} \psi + \eta  {\bar \psi} ]} \nonumber \\
\label{zfq}
\eea
Under the infinitesimal gauge transformation the quantum gluon field $Q_\mu^a$ transforms as
\bea
\delta Q_\mu^a = -gf^{abc}\omega^b Q_\mu^c + \partial_\mu \omega^a.
\label{lambda}
\eea
In the background field method of QCD the generating functional is given by \cite{thooft,abbott}
\bea
&& Z_{\rm background ~QCD}[A,J,\eta,{\bar \eta}]=\int [dQ] [d{\bar \psi}] [d \psi ] ~{\rm det}(\frac{\delta G^a(Q)}{\delta \omega^b}) \nonumber \\
&& e^{i\int d^4x [-\frac{1}{4}{F^a}_{\mu \nu}^2[A+Q] -\frac{1}{2 \alpha}
(G^a(Q))^2+{\bar \psi} \dslash [A+Q] \psi + J \cdot Q +{\bar \eta} \psi + \eta {\bar \psi} ]}
\label{zaqcd}
\eea
where $A_\mu^a$ is the background field. The gauge fixing term is given by
\bea
G^a(Q) =\partial_\mu Q^{\mu a} + gf^{abc} A_\mu^b Q^{\mu c}=D_\mu[A]Q^{\mu a}
\label{ga}
\eea
which depends on the background field $A_\mu^a$. When the background field $A_\mu^a(x)$ is pure gauge in QCD given by
\bea
T^aA_\mu^a = \frac{1}{ig} (\partial_\mu U) U^{-1},~~~~~~~~~~~~~~~~~~~~~~~U=e^{igT^a\omega^a(x)}
\label{pure}
\eea
we find from \cite{nayakgf}
\bea
&& Z_{\rm QCD}[J,\eta,{\bar \eta}] ~=e^{i\int d^4x J \cdot A}~\times~Z_{\rm background ~QCD}[A,J,\eta,{\bar \eta}]-~\int [dQ] [d{\bar \psi}] [d \psi ]~{\rm det}(\frac{\delta G_f^a(Q)}{\delta \omega^b}) \nonumber \\
&&~e^{i\int d^4x [-\frac{1}{4}{F^a}_{\mu \nu}^2[Q] -\frac{1}{2 \alpha} (G_f^a(Q))^2+{\bar \psi} \dslash [Q] \psi + J \cdot Q +{\bar \eta} \psi
+ \eta {\bar \psi} ]} ~[i \int d^4x [ J \cdot \delta Q
+ {\bar \eta} \delta \psi + \eta \delta {\bar \psi}+...]],
\label{final}
\eea
where
\bea
G_f^a(Q) =\partial_\mu Q^{\mu a} + gf^{abc} A_\mu^b Q^{\mu c} - \partial_\mu A^{\mu a}=D_\mu[A] Q^{\mu a} - \partial_\mu A^{\mu a}.
\label{gfa}
\eea
Changing the variable $Q \rightarrow Q-A$ in eq. (\ref{zaqcd}) we find
\bea
&& Z_{\rm background ~QCD}[A,J,\eta,{\bar \eta}]= e^{-i\int d^4x J \cdot A}~ \int [dQ] [d{\bar \psi}] [d \psi ] ~{\rm det}(\frac{\delta G_f^a(Q)}{\delta \omega^b}) \nonumber \\
&& e^{i\int d^4x [-\frac{1}{4}{F^a}_{\mu \nu}^2[Q] -\frac{1}{2 \alpha} (G_f^a(Q))^2+{\bar \psi} \dslash [Q] \psi + J \cdot Q + \eta {\bar \psi}
+{\bar \eta} \psi ]}.
\label{zaqcd1}
\eea
The fermion fields and the corresponding sources transform as follows
\bea
\psi' = U \psi,~~~~~~~~~{\bar \psi}' = {\bar \psi} U^{-1},~~~~~~~~~~~~~~\eta' = U \eta,~~~~~~~~~{\bar \eta}' = {\bar \eta} U^{-1}.
\label{ftq}
\eea
This gives
\bea
\delta (\eta {\bar \psi}) =\eta' {\bar \psi}'-\eta {\bar \psi}= 0 = \delta ({\bar \eta} \psi).
\label{etbps}
\eea
Using eq. (\ref{etbps}) in (\ref{final}) we find
\bea
&& Z_{\rm QCD}[J,\eta,{\bar \eta}] ~=e^{i\int d^4x J \cdot A}~\times~Z_{\rm background ~QCD}[A,J,\eta,{\bar \eta}]+~\int [dQ] [d{\bar \psi}] [d \psi ]~{\rm det}(\frac{\delta G_f^a(Q)}{\delta \omega^b}) \nonumber \\
&&~e^{i\int d^4x [-\frac{1}{4}{F^a}_{\mu \nu}^2[Q] -\frac{1}{2 \alpha} (G_f^a(Q))^2+{\bar \psi} \dslash [Q] \psi + J \cdot Q +{\bar \eta} \psi
+ \eta {\bar \psi} ]} ~[i \int d^4x [ -J \cdot \delta Q
+ \psi \delta {\bar \eta} + {\bar \psi} \delta \eta +...]].
\label{final1}
\eea
Hence from eqs. (\ref{final1}), (\ref{zaqcd1}) and (\ref{ftq}) we finally obtain
\bea
&& Z_{\rm QCD}[J,\eta,{\bar \eta}] ~=e^{i\int d^4x J \cdot A}~\times~Z_{\rm background ~QCD}[A,J,\eta',{\bar \eta}']-~\int [dQ] [d{\bar \psi}] [d \psi ]~{\rm det}(\frac{\delta G_f^a(Q)}{\delta \omega^b}) \nonumber \\
&&~e^{i\int d^4x [-\frac{1}{4}{F^a}_{\mu \nu}^2[Q] -\frac{1}{2 \alpha} (G_f^a(Q))^2+{\bar \psi} \dslash [Q] \psi + J \cdot Q +{\bar \eta} \psi
+  \eta {\bar \psi} ]} ~[i \int d^4x  J \cdot \delta Q+...].
\label{finalq}
\eea

The correlation function in the presence of the background field $A_\mu^a(x)$ is given by
\bea
\frac{\delta}{\delta {\eta}(x_2) }~\frac{\delta}{\delta {\bar \eta}(x_1) }
~Z_{\rm background ~QCD}[A,J,\eta,{\bar \eta}]|_{J=\eta={\bar \eta}=0}~~=~~<{\bar \psi}(x_2) {\psi}(x_1)>^A_{\rm background~QCD}.
\label{g1q}
\eea
The correlation function in QCD (without the background field) is given by
\bea
\frac{\delta}{\delta {\eta}(x_2) }~\frac{\delta}{\delta {\bar \eta}(x_1) }
~Z_{\rm QCD}[J,\eta,{\bar \eta}]|_{J=\eta={\bar \eta}=0}~~=~~<{\bar \psi}(x_2) {\psi}(x_1)>^{A=0}_{\rm QCD}.
\label{g2q}
\eea
Using eqs. (\ref{finalq}), (\ref{g1q}), (\ref{g2q}) and (\ref{ftq}) we find
\bea
<{\bar \psi}(x_2) U(x_2) U^{-1}(x_1) {\psi}(x_1)>^A_{\rm background~QCD}~=<{\bar \psi}(x_2) {\psi}(x_1)>^{A=0}_{\rm QCD}.
\label{fact1fa}
\eea
By taking appropriate color traces in eq. (\ref{fact1fa}) and by using eqs. (\ref{wilf1a}) and (\ref{wilf1}) we find
\bea
&& {\rm Tr}_{\rm color}~<{\bar \psi}(x_2) {\psi}(x_1)>^A_{\rm background~QCD}~=
 ~{\rm Tr}_{\rm color}~[{\cal P}~ {\rm exp}[-ig\int_{-\infty}^{0} d\lambda~ h \cdot {\cal A}^a(x^\mu_2 +h^\mu \lambda )T^a]] \nonumber \\
&&~\times ~[<{\bar \psi}(x_2) { \psi}(x_1)>^{A=0}_{\rm QCD}]~\times ~[{\bar {\cal P}}~ {\rm exp}[ig\int_{-\infty}^0 d\lambda~ h \cdot {\cal A}^b(x^\nu_1 +h^\nu \lambda )T^b]]
\label{fact1f}
\eea
which proves the factorization theorem by using background field method of QCD. All the ${\cal A}_\mu^a(x)$ dependences
(responsible for soft and collinear divergences) have been factored into the path ordered exponentials or Wilson lines.
This concludes the proof of factorization theorem in high energy QCD through symmetry considerations at the
lagrangian level.

\acknowledgements

This work was supported in part by Department of Energy under contracts DE-FG02-91ER40664,
DE-FG02-04ER41319 and DE-FG02-04ER41298.

\end{document}